\documentclass[%
preprint,
 amsmath,amssymb,
 aps,
]{revtex4-2}

\usepackage{graphicx}
\usepackage{dcolumn}
\usepackage{bm}
\usepackage{amsmath}
\usepackage{xcolor}

\usepackage{hyperref}
\hypersetup{
    colorlinks=true,
    urlcolor= blue,
    citecolor=blue,
linkcolor= blue}
\usepackage{xcolor}

\begin{document}

\preprint{APS/123-QED}

\title{Temperature dependence and limiting mechanisms of the upper critical field of FeSe thin films}

\author{M. Stanley\textsuperscript{1}}
\author{Y. Li\textsuperscript{1}}
\author{J. C. Palmstrom\textsuperscript{2}}
\author{J. L. Thompson\textsuperscript{3}}
\author{K. D. Halanayake\textsuperscript{3}}
\author{D. Reifsnyder-Hickey\textsuperscript{3}}
\author{R. D. McDonald\textsuperscript{2}}
\author{S. A. Crooker\textsuperscript{2}}
\author{N. Trivedi\textsuperscript{4}}
\author{N. Samarth\textsuperscript{1,5}}
\email{nsamarth@psu.edu}
\affiliation{\textsuperscript{1}Department of Physics, The Pennsylvania State University, University Park, Pennsylvania 16802, USA \\
\textsuperscript{2}National High Magnetic Field Laboratory, Los Alamos National Laboratory, Los Alamos, New Mexico 87545, USA\\
\textsuperscript{3}Department of Chemistry, The Pennsylvania State University, University Park, Pennsylvania 16802, USA \\
\textsuperscript{4}Department of Physics, The Ohio State University, Columbus, Ohio 43210, USA}
\affiliation{\textsuperscript{5}Department of Materials Science and Engineering, The Pennsylvania State University, University Park, Pennsylvania 16802, USA}

\begin{abstract}
We use magnetoresistance measurements at high magnetic field ($B \leq 65$ T) and low temperature ($T \geq 500$ mK) to gain fresh insights into the behavior of the upper critical field, $H_{c2}$, in superconducting ultrathin FeSe films of varying degrees of disorder, grown by molecular beam epitaxy on SrTiO$_3$. Measurements of $H_{c2}$ across samples with a widely varying superconducting critical temperature (1.2 K~$ \leq T_c \leq 21$~K) generically show similar qualitative temperature dependence. We analyze the temperature dependence of $H_{c2}$ in the context of Werthamer-Helfand-Hohenberg (WHH) theory. The analysis yields parameters that indicate a strong Pauli paramagnetic pair-breaking mechanism which is also reflected by pseudo-isotropic superconductivity in the limit of zero temperature. In the lower $T_c$ samples, we observe a spin-orbit scattering driven enhancement of $H_{c2}$ above the strongly-coupled Pauli paramagnetic limit. We also observe clear deviations from WHH theory at low temperature, regardless of $T_c$. We attribute this to the multi-band superconductivity of FeSe and possibly to the emergence of a low temperature, high field superconducting phase.
\end{abstract}

\maketitle


\section{\label{sec:level1}Introduction}

Understanding superconductivity in the structurally and stoichiometrically simple material system FeSe has been the target of many experimental efforts since its initial discovery in tetragonal phase single crystals with $T_{c} \approx 9 - 13.5$ K \cite{Hsu2008,Mizuguchi2008}. The superconducting behavior of FeSe thin films in the 2D limit has received a particular surge in attention in recent years after reports of significantly enhanced superconducting gap closing critical temperatures up to 65 K when interfaced with SrTiO$_3$ (STO) substrates, as measured by \emph{in vacuo} angle-resolved photoemission spectroscopy (ARPES) and scanning tunneling spectroscopy (STS) \cite{Wang2012, He2013, Zhang2014, Liu2012, Lee2014, Tan2013}. {\it In vacuo} electrical transport measurements on pristine (uncapped) 2D FeSe films grown on STO show reproducible zero resistance at $T_c$ as high as $\sim 30$~K \cite{Faeth_PhysRevX.11.021054, li2021,Ide_PhysRevMaterials.6.124801}, significantly higher than in FeSe bulk crystals and consistent with the large superconducting gap seen in ARPES and STS. Despite many experimental studies of superconductivity in 2D FeSe/STO \cite{Huang_doi:10.1146/annurev-conmatphys-031016-025242,Kreisel_2020}, there are still opportunities to improve our understanding of the superconductivity of this material system. One pathway is to study the upper critical field, $H_{c2}$, a fundamental parameter of superconductivity, which represents the strength at which an externally applied magnetic field will break down the Cooper pairs of a superconductor.

Pair-breaking at $H_{c2}$ is generally considered to have two origins: the polarization of the spins of the constituent electrons of Cooper pairs via Pauli paramagnetism, and the orbital effect when cores of Abrikosov vortices start to overlap. It is an essential parameter for understanding numerous superconducting properties of a material, including the coherence length, the dimensionality of superconductivity, and anisotropy. Systematic measurements of $H_{c2}$ can also provide insight into pair-breaking mechanisms and their relative strengths. Due to its relatively low $T_{c}$, bulk, single crystal FeSe has been an ideal subject for several high magnetic field measurements aimed at establishing the overall behavior and dependencies of $H_{c2}$ in the limit of zero temperature and at various pressures and sample thicknesses \cite{Lei2012, Vedeneev2013, Kang2016_1, Farrar2020}. However, similarly rigorous studies of $H_{c2}$ have been technically challenging in ultrathin films of FeSe grown on STO because the greatly enhanced values of $T_{c}$ are accompanied by extremely large values of
$H_{c2}$ \cite{Zhou2016}. To our knowledge, a detailed analysis of the temperature dependence of $H_{c2}$ for FeSe/STO has not yet been reported. In this paper, we use high magnetic field ($B \leq 65$ T) magnetoresistance (MR) measurements to map out the variation of $H_{c2}$ as a function of temperature and disorder in four ultrathin FeSe films grown by molecular beam epitaxy (MBE) on STO. Our observations provide important insights into the unique interfacial superconductivity in this material system.    

\section{Experiment}
To produce FeSe thin films capable of achieving high temperature superconductivity, $T_{c,0} \geq$ 20 K, we first treated commercially obtained (Shinkosha) STO (001) substrates with a chemical etching process which included 45 minutes in deionized water at $90 ^{\circ}$C and another 45 minutes in a 25\% HCl solution at room temperature. Substrates were then annealed under a flow of oxygen in a tube furnace for 3 hours at 980$ ^{\circ}$C, rendering atomically flat, TiO$_2$-terminated surfaces. Substrates were additionally outgassed \emph{in vacuo} at 500$ ^{\circ}$C for 2 hours prior to growth.\par 
We deposited the FeSe thin films using a Scienta Omicron EVO-50 MBE system, with a base pressure of 10$^{-11}$ Torr, using high purity Fe (99.99\%) and Se (99.999\%) elemental sources simultaneously evaporated from thermal Knudsen effusion cells. We maintained the substrate temperature at roughly $350 ^{\circ}$ C, as measured by an infrared pyrometer, and used Fe and Se fluxes that yielded a deposition rate of roughly 1 layer every 6.5 minutes. This yielded films of around 3 unit cells (U.C.) thickness (1.65 nm) after a typical growth period of 20 minutes (Fig. 2). A post-growth anneal was performed at a substrate temperature between 525$ ^{\circ}$C and 550$ ^{\circ}$C for a duration varying between 2 to 3 hours. This annealing process is expected to induce Se vacancies, allowing for electron charge transfer into the film from the STO, and is a critical step in the realization of high temperature superconductivity in this system \cite{Zhang2014_2, Berlijn2014}. We grew four samples, A, B, C, and D, with varying post-growth annealing conditions, yielding a large variation in $T_C$. Prior to their removal from the UHV chamber, samples were capped with approximately 10 U.C.s (5-6 nm) of crystalline FeTe, grown at 350$ ^{\circ}$C for surface passivation, and 10 nm of amorphous Te deposited at room temperature for additional protection from oxidation; we find that this capping protocol provides consistent results when studying the superconducting properties of FeSe/STO films {\it ex situ}\cite{li2021}. Grown on 5 x 5 mm$^{2}$ pieces of STO held in place with clips of tantalum foil, the typical coverage of films in this study was around 10 mm$^{2}$. \par

We then prepared samples for {\it ex situ} electrical transport measurements by mechanically etching six-terminal Hall bar patterns with effective areas of $0.5 \times 1.0 \textrm{mm} ^{2}$. Electrical contacts were formed with pressed indium dots. Samples were initially evaluated in a Quantum Design 14 T Physical Property Measurement System in order to identify candidate samples representing different levels of disorder. Finally, we measured field-dependent longitudinal resistivities $\rho_{xx}(H^{ab})$ and $\rho_{xx}(H^{c})$ in pulsed magnetic fields up to 65 T applied in plane and out of plane, respectively. We focus on four particular samples (A, B, C, and D) of similar thickness but with widely varying zero resistance critical temperatures (Fig. 1). \par 

Samples were mounted on a transport probe and loaded into a cryostat with a double-walled vacuum jacket, which in turn was immersed in a bath of liquid $^{4}$He. We achieved temperatures down to 500 mK by condensing $^{3}$He into the cryostat and subsequent pumping. The capacitor-driven 65 T pulsed magnet produced 80 ms pulses, with a rise time of 8 ms. Eddy currents induced by this rapidly changing magnetic field, or \emph{dB/dt}, result in self-heating effects which can become non-negligible at very low temperatures. Thermal stability was maximized by measuring in liquid $^{4}$He for temperatures between 1.2 K and 4K. Additionally, only data gathered during the down-sweep of the magnetic field pulse, where the \emph{dB/dt} was much smaller, was considered during our analysis. 

After performing all other characterization, the samples were analyzed by annular dark-field scanning transmission electron microscopy (ADF-STEM) imaging. Samples were first prepared by depositing ~15 nm of amorphous carbon in a Leica sputter coater, and then they were transferred into an FEI Scios 2 dual-beam scanning electron microscopy focused ion beam (SEM-FIB) for the creation of electron-transparent lamellae for STEM imaging. The lamellae were extracted from the regions that had previously been patterned for electrical transport measurements, and they were subsequently thinned to electron transparency using FIB accelerating voltages of 30 kV and 5 kV, with a final cleaning step at 2 kV. ADF-STEM imaging and energy-dispersive X-ray spectroscopy (EDX) elemental analysis were performed on a dual-spherical aberration-corrected FEI Titan3 G2 60-300 S/TEM at an accelerating voltage of 300 kV with a convergence angle of 25.2 mrad. Images were collected by a high-angle ADF detector with collection angles of 42-244 mrad.

\section{Results and Discussion}
The temperature dependent resistivity for all 4 samples (A, B, C, D) is shown in Fig. 1. Samples A, B, and C show similar behavior in the normal state, where $\rho_{xx} (T)$ is almost independent of temperature before reaching the superconducting transition at their respective values of $T_c$. The qualitative behavior of $\rho_{xx}(T)$ in sample D resembles the metallic behavior seen in bulk crystals of FeSe, monotonically decreasing with decreasing temperature, albeit with a smaller residual resistivity ratio (RRR) $\sim 2$ than observed in high quality bulk crystals \cite{Hsu2008, Mizuguchi2008, Bohmer2016}. Sample D exhibits the highest zero resistance critical temperature, as well as the highest RRR, indicating that it is the least disordered of the thin film samples we studied. These sample-specific parameters and others are outlined in Table 1. 
\par
It is interesting to note that sample A with the highest disorder shows a small increase in resistivity just before going superconducting. The increase in resistivity in the more disordered sample is consistent with emergent granularity. In the normal state, the grains or patches become locally superconducting but the different patches are not Josephson coupled. As the temperature is reduced, the Josephson coupling increases and the patches get coupled into a globally coherent state~\cite{Ghosal1,Ghosal2,Bouadim}.  
\par
The onset superconducting critical temperature $T_{c}^{onset}$ and normal state resistivity $\rho_{n}^{onset}$ for each sample are determined from the intersections of two extrapolated lines. The first is drawn through the steepest portion of $\rho_{xx}(T)$ within the superconducting transition and the second through $\rho_{xx}(T)$ in the normal state just above the superconducting transition (Fig. 1 inset). 
\par  

Since the four samples presented in this study were grown under similar growth conditions, we performed scanning transmission electron microscopy (STEM) measurements in an effort to understand the large variation in their superconducting critical temperatures from a microscopic point of view. The cross-sectional high-angle annular dark-field (HAADF) images in Fig. 2 compare samples A and D, the lowest and highest $T_c$ samples, respectively. Sample D shows the expected 3 U.C. FeSe (001) epitaxial layer, and clearly defined FeTe and Te capping layers. While the TEM measurements potentially indicate a greater level of disorder in sample A, we note that the lack of clarity in the image (Fig. 2(a)) most likely arises from some form of contamination (perhaps due to In contacts during sample preparation for TEM measurements); it cannot be directly associated with crystallographic disorder. We note that these STEM measurements were carried out on the actual sample used in high field measurements after the transport experiments were performed. More detailed STEM data for these samples and for sample B, C is available in the Supplementary Material \cite{Stanley_supp}. 

\begin{table}
\caption{\label{tab:table3}Parameters related to the superconducting transition of each sample extracted from Fig. 1. The temperature at which 90\% of the sample resistance at $T_{c, \textrm{onset}}$ is reached defines $T_{c, 90}$, and $T_{c, 0}$ is the point at which the resistance is less than 5\% of $T_{c, \textrm{onset}}$ effectively close to zero. The RRR is calculated as $\rho (293K) / \rho (T_{c, \textrm{onset}})$. }
\begin{ruledtabular}
\begin{tabular}{|c|c|c|c|c|}
 
 Sample & $T_{c, \textrm{onset}}$ [K[ \: & $T_{c, 90}$ [K]\:& $T_{c, 0}$ [K]& RRR \:\:\:\:
\\ \hline
 A   &11.0 \:  &10.3 \:\:& 1.2 \:\:& 1.1 \:\:\:\:  \\
 B   &12.6 \:  &11.8 \:\:& 8.5 \:\:& 1.4  \:\:\:\:  \\
 C   &24.0 \:  &22.0 \:\:& 14.0 \:\:& 1.2  \:\:\:\:  \\
 D   &32.0 \:  &27.2 \:\:& 21.0 \:\:& 2.1   \:\:\:\: \\

\end{tabular}
\end{ruledtabular}
\end{table}

The results of MR measurements $\rho_{xx}(\mu_{0}H)$ in pulsed high magnetic fields are shown in Fig. 3 for sample A in both $H \parallel ab$ and $H \parallel c$ geometries. The temperature dependence of $H_{c2}$ is extracted from these plots by identifying the abscissa corresponding to 10\% ($\rho_{n}^{10}$), 50\% ($\rho_{n}^{50}$), and 90\% ($\rho_{n}^{90}$) of the normal state resistivity along each isotherm. These positions correspond to three separate but self-consistent definitions of $H_{c2}$. Similar data for samples B, C, and D are shown in the Supplementary Material \cite{Stanley_supp}.
\par
For samples A and B (which have the lowest $T_c$), we can determine $H_{c2}(T)$ down to the $^4$He base temperature in both field orientations. For samples C and D (which have higher $T_c$), the available data in the $H \parallel ab$ geometry was restricted to temperatures near the superconducting transition because $H_{c2}(T)$ exceeds the available field strength (65 T). This observation is consistent with the fact that orbital-limiting effects in the $H \parallel ab$ configuration should be suppressed for such a highly anisotropic few-layer system where vortex formation is confined to a relatively small cross sectional area. \\
\indent Measured temperatures, in general, were selected in order to comprehensively sample the superconducting and normal states, as well as the entire breadth of the superconducting transition region in each film. The resulting values of $H_{c2}(T)$ for sample A are compiled in Fig. 4 as an example. We found that the qualitative behavior of $H_{c2}(T)$ was not heavily dependent on the resistivity criteria used to define it, possibly indicating minimal field-induced broadening of the superconducting transition in these samples. We therefore proceed strictly with the $\rho_{n}^{90}$ criteria for defining $H_{c2}(T)$ due to the abundance of data points it provides, without loss of generality. 

We analyze our data using Werthamer-Helfand-Hohenberg (WHH) theory which provides a prediction of the temperature dependence of $H_{c2}(T)$ for type-II superconductors, and can account for the effects of both orbital and paramagnetic limitation \cite{Helfand1964, Helfand1966, Werthamer1966}. In the dirty limit, where the electron mean free path $l$ is much smaller than the superconducting coherence length $\xi$, $H_{c2}$(T) is determined when the following condition is met \cite{Solenov2017}: 
{\small
\begin{align}
\centering
0 = \ln (t) &  - \psi(\frac{1}{2}) + (1 + \frac{i\lambda_{so}}{\sqrt{4\alpha^{2} h^{2} - \lambda_{so}^{2}}}) \\ 
& \times \psi([2h + \lambda_{so} +2t + i\sqrt{4\alpha^{2} h^{2} - \lambda_{so}^{2}}]/4t)/2  
+ c.c. \nonumber
\end{align}
}%
\noindent
Here, $\hbar = k_{b} = 1$, $\psi (x)$ is the complex digamma function, $t = T/T_c$ is the normalized temperature, $h = eH_{c2}\nu_{F}^{2}\tau / 3\pi T_{c}$ is the reduced magnetic field, and the Maki parameter $\alpha$ and spin-orbit coupling parameter $\lambda_{so}$ are defined as:
\begin{align}
    \alpha = 3/2m\nu_{F}^{2}\tau\\
    \lambda_{so} = 1/3\pi T_{c}\tau_{so} \nonumber
\end{align}

\noindent
Last, we define $\overline{H} \equiv \alpha h = eH_{c2}/2\pi T_{c}m$ as another dimensionless representation of the upper critical magnetic field convenient for plotting. \par
Special cases of the upper critical field arise first in the absence of both Pauli paramagnetic limitation and spin-orbit interaction, when $\alpha = \lambda_{so} = 0$, where the entirely orbital-limited $H_{c2}$ is determined to be \cite{Werthamer1966}:
\begin{equation}
    \mu_0 H_{c2}^{orb}(0) = -0.693T_{c}\left(-\frac{d\mu_{0}H}{dT}\right)\vert_{T=T_{c}}
\end{equation}
Second, if $H_{c2}$ is instead entirely Pauli-limited, superconductivity is quenched when the Zeeman energy exceeds the superconducting condensation energy. In the case of a weakly-coupled BCS superconductor, this corresponds to the Chandrasekhar-Clogston limit \cite{Chandrasekhar1962, Clogston1962}:
\begin{equation}
    \mu_{0}H_{c2}^{P}(0) = 1.86\: T_{c}
\end{equation}
This assumes that $2\Delta_0 = 3.5k_{b}T_{c}$.
\par 
The Maki parameter (Eq. 2) describes the relative strengths of the Pauli paramagnetic effect and the orbital limiting field when these two mechanisms have comparable effects in the pair-breaking process \cite{Maki1966}. In this case, it can be written as $\alpha = \sqrt{2} H_{c2}^{orb}/H_{c2}^{P}$. Experimentally, the Maki parameter can be interpreted from the slope of $H_{c2}(T)$ in the vicinity of $T_{c}$ using this relation and Eqns. 3 and 4:
\begin{equation}
    \alpha \approx 0.53 \left(-\frac{d\mu_{0}H}{dT}\right)\vert_{T=T_{c}}
\end{equation}

We numerically solve Eq. 1 using $\alpha$ and $\lambda_{so}$ as fitting parameters for the experimental $\overline{H}_{c2}^{ab}(t)$ and $\overline{H}_{c2}^{c}(t)$ data of each sample. The numerical results are summarized in Table 2 along with the values from theoretical predictions.  The magnitude of the spin-orbit coupling parameter, $\lambda_{so}$, is indicative of the spin-flip scattering strength in the sample, and a non-zero value of $\lambda_{so}$ incorporated into the WHH model enhances the predicted values of $H_{c2}(T)$, especially in the zero temperature limit. Physically, this corresponds to a suppression of the Pauli paramagnetic effect through a reduction of the Zeeman energy. 

We now discuss the WHH fits of the temperature dependence of the upper critical field for all four samples (Fig. 5). Here, we assume $\lambda_{el-ph} = 0.3$ \cite{Li2014}. Each plot includes additional WHH curves constructed using the non-zero Maki parameter calculated while fitting experimental data but without the contribution of spin-orbit scattering to highlight the necessity of a non-zero $\lambda_{so}$. Additionally, for sample A, we have calculated five WHH curves with $\alpha$ from 1 to 5 and $\lambda_{so} = 0$ (see Supplementary Material \cite{Stanley_supp}). All these plots reveal a clear necessity for including a non-zero $\lambda_{so}$ in order to accurately fit the experimental data for both the parallel and perpendicular field geometries, reflecting the significant role of spin-orbit coupling in FeSe \cite{Kang2016_1, Ma2017}.   
\par

In the $H \parallel ab$ geometry, we observe very large Maki parameters in all four samples. Physically, this suggests weak orbital limitation, and conversely, stronger Pauli-limiting effects in this configuration. This result is consistent with the picture of vortex formation and flow being hindered by the relatively small cross-sectional area of the $\sim$1.5 nm thick films during the superconducting-normal state transition. The fits in Fig. 5 (a)-(d) also show a large deviation between the WHH fit of the $H \parallel ab$ data and the theoretical fully-orbital-limited curves ($\alpha = 0, \lambda_{so} = 0$), especially towards zero temperature, as another indication of weak orbital-limiting effects. Finally, to this point, we note that the extracted Maki parameters are greater than those calculated under the assumption of comparable limiting mechanisms (Table 2) by nearly a factor of two for each sample. This further suggests the significance of Pauli-limiting effects here. 
\par 
The values of the Maki and spin-orbit scattering parameters derived from our WHH analysis are larger than observed in single crystal FeSe samples. Reported values in bulk FeSe crystals include $\alpha = 0.82$ ($H \parallel ab$) and $\alpha = 0.37$ ($H \parallel c$) \cite{Kang2016_1}, and $\alpha = 2.1$ ($H \parallel ab$) \cite{Audouard2015}, both without contribution from a non-zero $\lambda_{so}$. In thin flakes of FeSe, values of $\alpha = 2.4$ and $\alpha = 4.15$ have been observed for sample thicknesses of $t = 54, 100$ nm, and $t = 24$ nm, respectively \cite{Farrar2020}. Complete descriptions of the data in this case of thin flakes required $\lambda_{so} = 0.2 - 0.35$. Results obtained from the work presented here ($\alpha = 8.7 - 9.6$ for H $\parallel$ ab and $\alpha = 1.1 - 3.2$ for H $\parallel$ c) on $t \approx 1.5$ nm thin films of FeSe are consistent with a trend of increased Maki parameters, as well as the inclusion of spin-orbit scattering into the WHH model to effectively describe the data, in the 2D limit. 

\begin{table*}
\scriptsize
\caption{\label{tab:table2}Overview of various properties and superconducting parameters for each film discussed here. The columns are, in order, the sample name, the Maki parameters extracted from the WHH fitting for both $H \parallel ab$ and $H \parallel c$ geometries, the corresponding spin-orbit scattering parameters, the zero-temperature $H_{c2}(T)$ predicted by WHH theory with best fit parameters,  the theoretically predicted orbital-limited $H_{c2}$ at zero temperature, the weakly-coupled Pauli-paramagnetic limit of $H_{c2}$ at zero temperature.}
\begin{ruledtabular}
\begin{tabular}{|c|c|c|c|c|c|c|c|}
 
 Sample & $\alpha \: (\parallel ab , \parallel c)$ & $\lambda_{so} \: (\parallel ab , \parallel c)$ & $H_{c2}^{ab}(0)$ & $H_{c2}^{c}(0)$ & $H_{c2}^{orb}(0) \parallel ab$ & $H_{c2}^{orb}(0) \parallel c$ 
& $H_{c2}^{P}(0)$  \\ \hline
 A   &9.6, 2.3&1.8, 1.7&34.2 T& 21.8 T&71.1 T & 26.1 T &19.2 T\\
 B   &8.7, 3.2&1.6, 2.3&37.1 T &34.6 T&83.5 T & 34.8 T &21.9 T \\
 C   &8.9, 2.5&1.2, 0.5&63.1 T &38.5 T&150.3 T & 67.6 T &40.9 T \\
 D   &9.4, 1.1&2.1, 1.5&94.2 T & 37.2 T&204.9 T & 32.1 T &50.6 T \\

\end{tabular}
\end{ruledtabular}
\end{table*}

\par
When the magnetic field is aligned perpendicular to the plane of the film, orbital effects are expected to play a more significant role, and this is demonstrated in part by the WHH fits of the H $\parallel$ c data in Fig. 5 (a-d). At higher temperatures, closer towards $T_{c}$, there is a clear reduction in the values of $H_{c2}$ compared to those from the H $\parallel$ ab measurements at comparable temperatures. Additionally, the calculated fit for samples A and B closely follow the purely orbital-limited curves ($\alpha = 0, \lambda_{so} = 0$). We also observe a clear deviation from the prediction of WHH as $H_{c2}^{c}$ increases almost linearly, upwards, in the limit of zero temperature in all four samples. Such upward curvature have been demonstrated in the '1111' \cite{Hunte2008, Jaroszynski2008} and '122' \cite{Baily2009, Yuan2009} systems, and has been discussed for the stoichiometric '11' Fe(Te,Se) system as well \cite{Fang2010,Khim2010}. This deviation from the single-band WHH theory is typically recognized as an effect of multi-band superconductivity \cite{Gurevich2003}.

\par  
In samples A and B, the low temperature $(T/T_{c} < 0.2)$ behavior of $H_{c2}^{ab}$ exhibits a slight upward curvature which deviates from the WHH prediction, even with the inclusion of a non-zero $\lambda_{so}$. Distinct high-field induced superconducting phases at low temperatures in single crystal FeSe samples have been reported \cite{Kasahara2014} and have been attributed to a possible unconventional Fulde-Ferrell-Larkin-Ovchinnikov (FFLO) state, even in the presence of disorder \cite{Zhou2021}. The FFLO state features spatial modulation of the superconducting order parameter and Cooper pairs of finite total momentum $(k\uparrow,-k+q \downarrow)$, compared to the conventional zero-momentum pairing $(k\uparrow,-k\downarrow)$. At high fields, Zeeman splitting plays a more significant role in quenching the superconducting phase, so a prerequisite for realization of the FFLO state is a large (greater than one) Maki parameter, indicating a smaller paramagnetically-limited field by the $\alpha = \sqrt{2} H_{c2}^{orb}/H_{c2}^{P}$ relation. This is indeed a feature in all four samples measured here. However, extremely large field strengths, much higher than 65 T, would be required to sufficiently survey the low temperature regime of higher-$T_c$ samples, such as C and D, for a more complete investigation of this unique state in the FeSe/STO system.

Next, we address the overall temperature dependence of the upper critical field anisotropy, quantified by the anisotropy parameter $\gamma(T)=H_{c2}^{ab}(T)/H_{c2}^{c}(T)$. This is shown in Fig. 6 and qualitatively agrees with that of other Fe-based superconductors, including single crystal FeSe \cite{Vedeneev2013}. In the limit of zero temperature, nearly isotropic superconductivity in other Fe-based superconducting systems, such as the ‘11’ and ‘122’ classes of materials, is commonly observed despite the quasi-2D nature of their electronic structures \cite{Zhang2011}. In the ultrathin limit, a significant increase in $\gamma(T)$ could rightfully be expected as the superconductivity further inherits the anisotropic character of the increasingly two-dimensional Fermi surface, but this behavior only seems to be observed in the neighborhood of the superconducting critical temperature \cite{Farrar2020}. Our four samples demonstrate varying degrees of enhanced anisotropy closer to their respective values of $T_{c}$, ranging from  $\gamma(t \approx 0.8) = 2.0 -  4.1$, as well as a monotonic decrease towards $\gamma = 1$ in the limit of zero temperature (Fig. 6). Overall, the orbital constraints on superconductivity in FeSe likely describe the larger anisotropy as the temperature approaches $T_{c}$: in this regime, the Maki parameter, and therefore the slope of $H_{c2}(T)$, is larger in the $H \parallel ab$ direction. In contrast, the approach to isotropic behavior of $H_{c2}(0)$ at lower temperatures can be attributed to the dominance of the Pauli-paramagnetic limit, so that $H_{c2}(0)$ is not affected by orientation of the field.
\par
Finally, we estimate the value of the zero-temperature coherence length, $\xi (0)$ from the extracted $H_{c2}(T)$ data using the 2D Ginzburg-Landau formula $\mu_0 H_{c2}^{\perp}(T) = \Phi_0 / (2 \pi \xi_{GL}(0)^2) \times (1-T/T_c)$ where $\Phi_0 = 2.07 \times 10^{-15}$ Wb is the superconducting magnetic flux quantum \cite{saito2015}. Fitting this equation to the data in the vicinity of $T_c$, we find that $\xi_{GL}(0) =$ 3.2 nm, 3.5 nm, 3.3 nm, and 4.8 nm  for samples A, B, C, and D, respectively.

 \section{Conclusion}

    In summary, we have studied the superconducting transition of four MBE-grown FeSe/STO thin films of varying disorder by determining their resistive upper critical fields using high magnetic field pulses up to 65 T applied in both parallel ($H \parallel ab$) and perpendicular ($H \parallel c$) directions. The experimental behavior of $H_{c2}(T)$, defined by the $\rho_n^{90}$ criterion, is in close agreement with WHH theory down to $T/T_{c} \approx 0.25$ when the effects of both orbital and Pauli paramagnetic limits are considered. Although our TEM measurements broadly show a correlation between structural disorder and $T_{c}$, the parameters extracted from the WHH fits are surprisingly insensitive to both the level of disorder and $T_{c}$. Prior HRTEM studies of the FeSe/STO interfacial structure have indicated that $T_{c}$ is possibly determined by the reconstruction of the FeSe Fermi surface resulting from excess Ti in a quasi-2D Ti$_{1-x}$O$_2$ interfacial layer \cite{Sims_PhysRevB.100.144103}. This has not been carefully examined in our samples and warrants further microscopy studies. Low temperature deviations of the $H_{c2}^{c}$ data from the theory are observed in each sample and may result from the effects of multi-band superconductivity. Similar deviations below $T/T_{c} \approx 0.2$ in the $H_{c2}^{ab}$ of samples A and B may indicate a separate high-field superconducting phase akin to the FFLO state. This warrants further investigation of higher $T_{c}$ samples, such as samples C and D, in fields with strengths exceeding 65 T to effectively study their detailed $H_{c2}(T)$ behavior in a low temperature regime. In future work we will further explore the combined role of spin-orbit coupling and disorder in a multiband superconductor such as FeSe.   
    

\begin{acknowledgments}
This work was supported by the Penn State 2DCC-MIP under NSF Grant No. DMR-2039351. Work at the National High Magnetic Field Lab was supported by NSF DMR-1644779, the State of Florida, and the U.S. Department of Energy and the research at Ohio State was supported by NSF Grant No. DMR2138905.

\end{acknowledgments}

%

\clearpage

\begin{figure}
    \centering
    \includegraphics[scale =1.0]{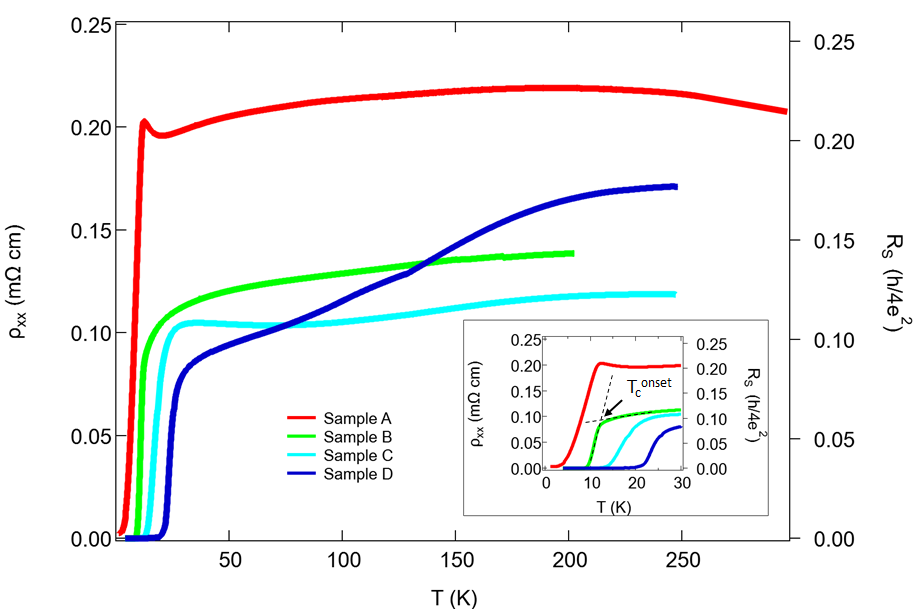}
    \caption{Temperature dependence of the resistivity $\rho_{xx}$ in samples A-D at zero magnetic field. The inset shows the superconducting transitions in greater detail over a more limited temperature range and the determination of $T_{c}^{onset}$. The sheet resistance is given in terms of the resistance quantum of a Cooper pair $h/4e^2$ = 6.45 k$\Omega$.} 
    \label{fig:Resistivity}
\end{figure}

\newpage

\begin{figure}
    \centering
    \includegraphics[scale=0.35]{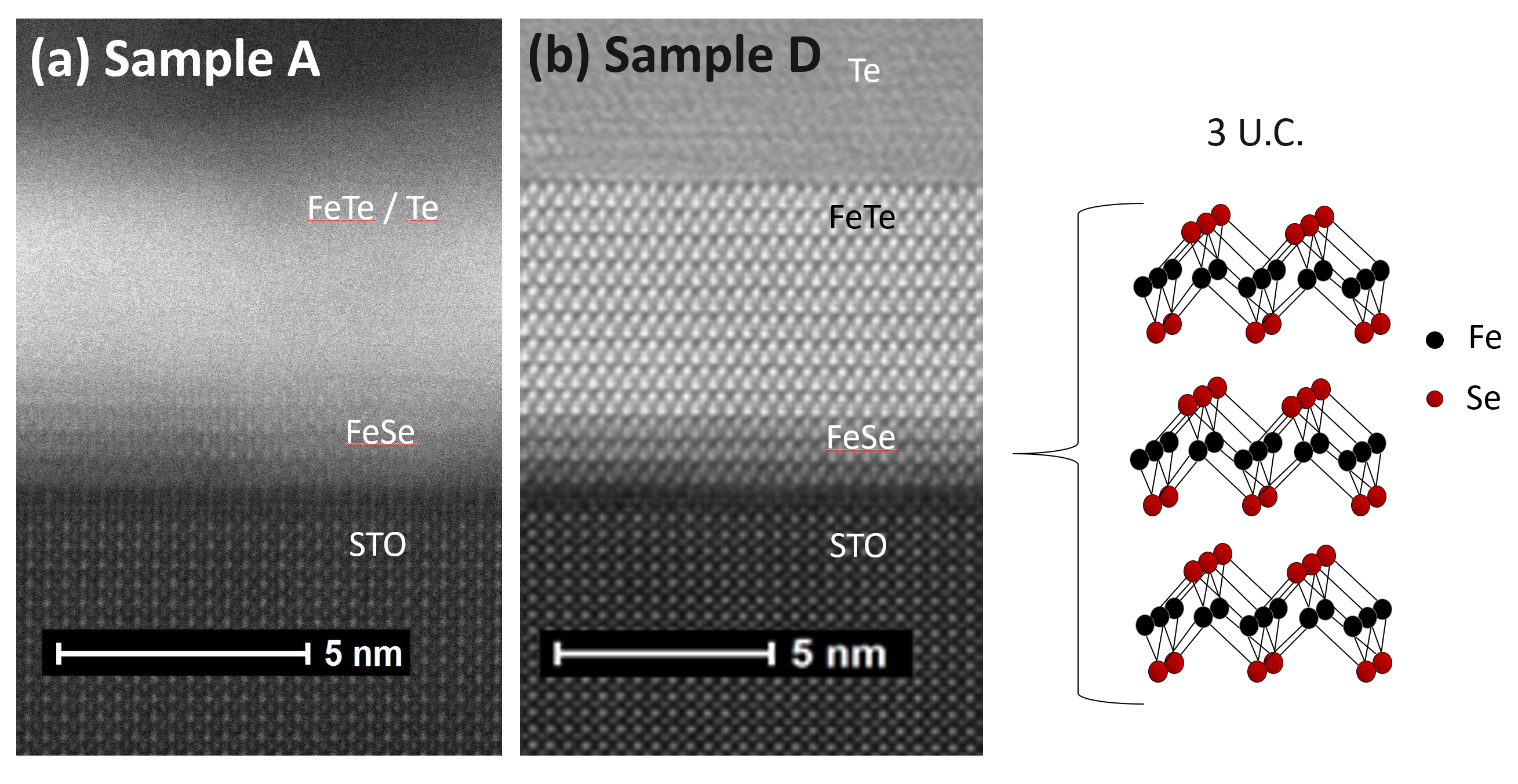}
    \caption{Cross-sectional HAADF-TEM images of the lowest and highest $T_c$ samples, A and D, respectively. Sample D clearly demonstrates a 3 unit-cell FeSe (001) epitaxial layer on STO (001), along with the expected crystalline FeTe and amorphous Te capping layers, while clear images of the grown layers in sample A could not be achieved.}
    \label{fig:crystal}
\end{figure}

\newpage 
\begin{figure}
    \centering
    \includegraphics[scale=0.5]{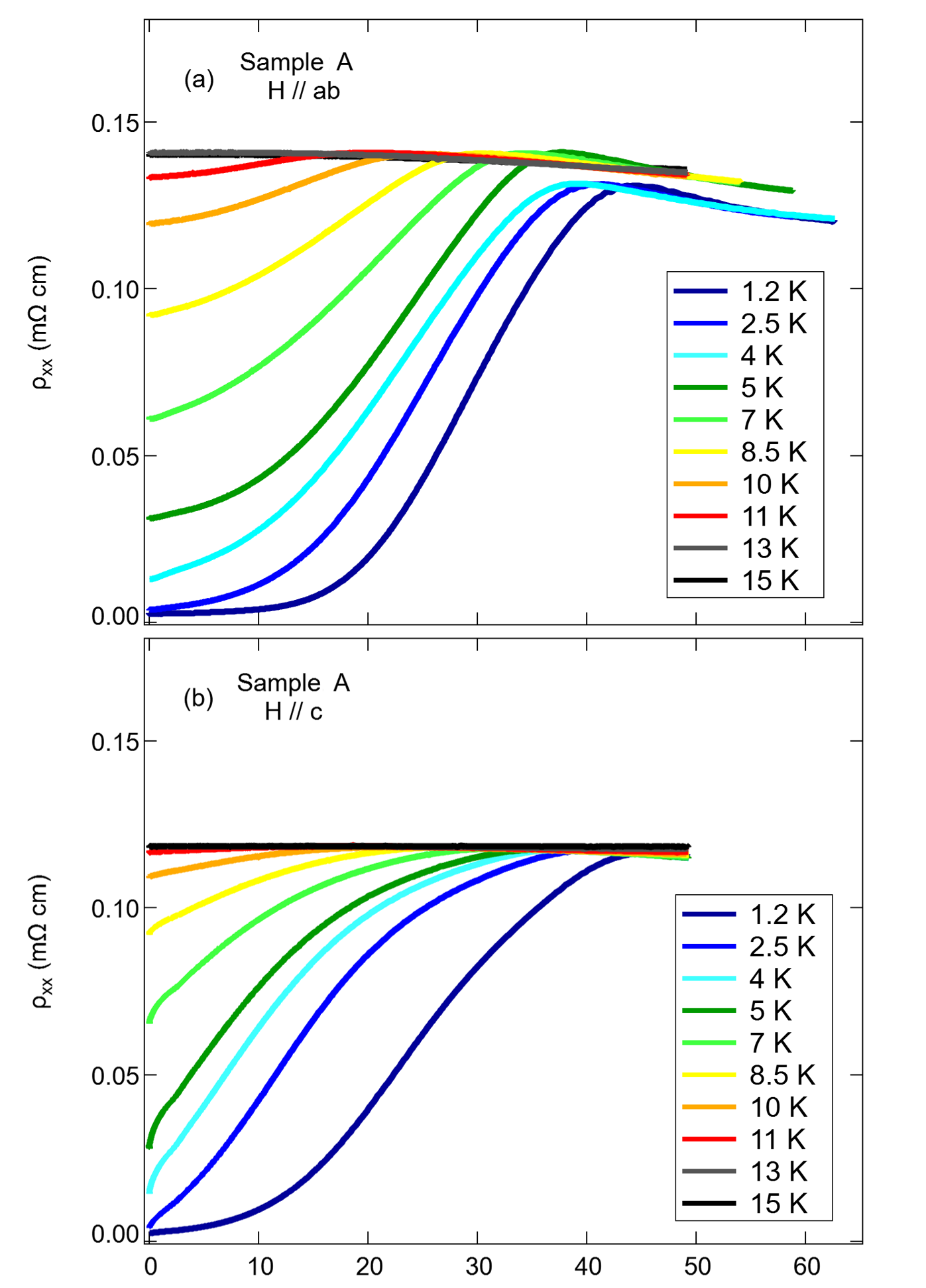}
    \caption{MR of sample A at various temperatures in the range $1.2$~K $\leq T \leq 15 $~K with magnetic field (a) in-plane ($H \parallel ab$) and (b) out-of-plane ($H \parallel c$).}
    \label{fig:Magnetoresistance}
\end{figure}

\newpage
\begin{figure}
    \centering
    \includegraphics[scale = 0.17]{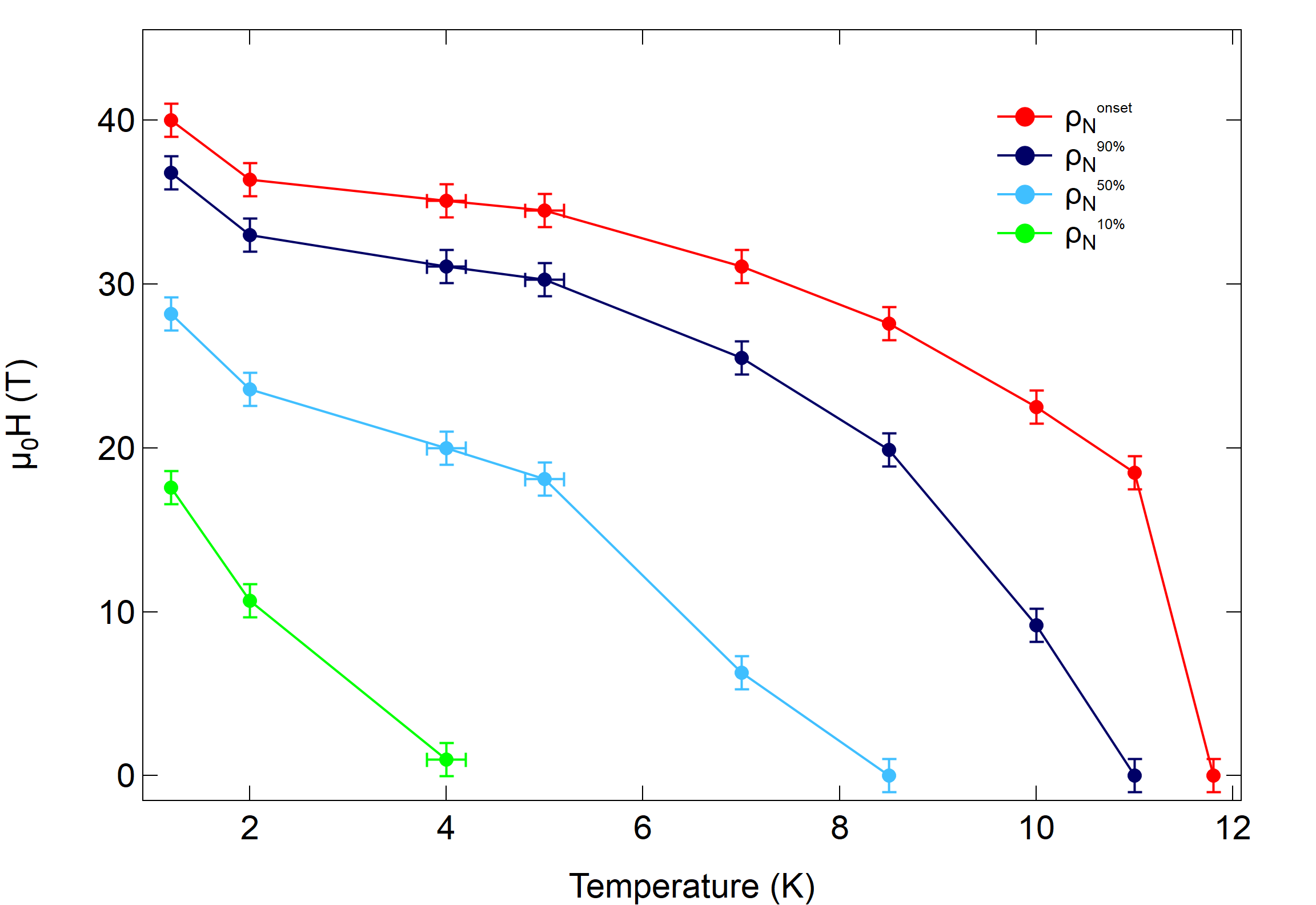}
    \caption{Temperature dependence of the upper critical field in the parallel field configuration $H \parallel ab$ of sample A using the $\rho_{n}^{10}$, $\rho_{n}^{50}$, $\rho_{n}^{90}$, and $\rho_{n}^{onset}$ criteria. The overall qualitative behavior of each curve is similar. Lines are a guide to the eye.}
    \label{fig:SampleA_Hc2}
\end{figure}

\newpage
 \begin{figure*}
      \centering
      \includegraphics[width=13cm,height=18cm]{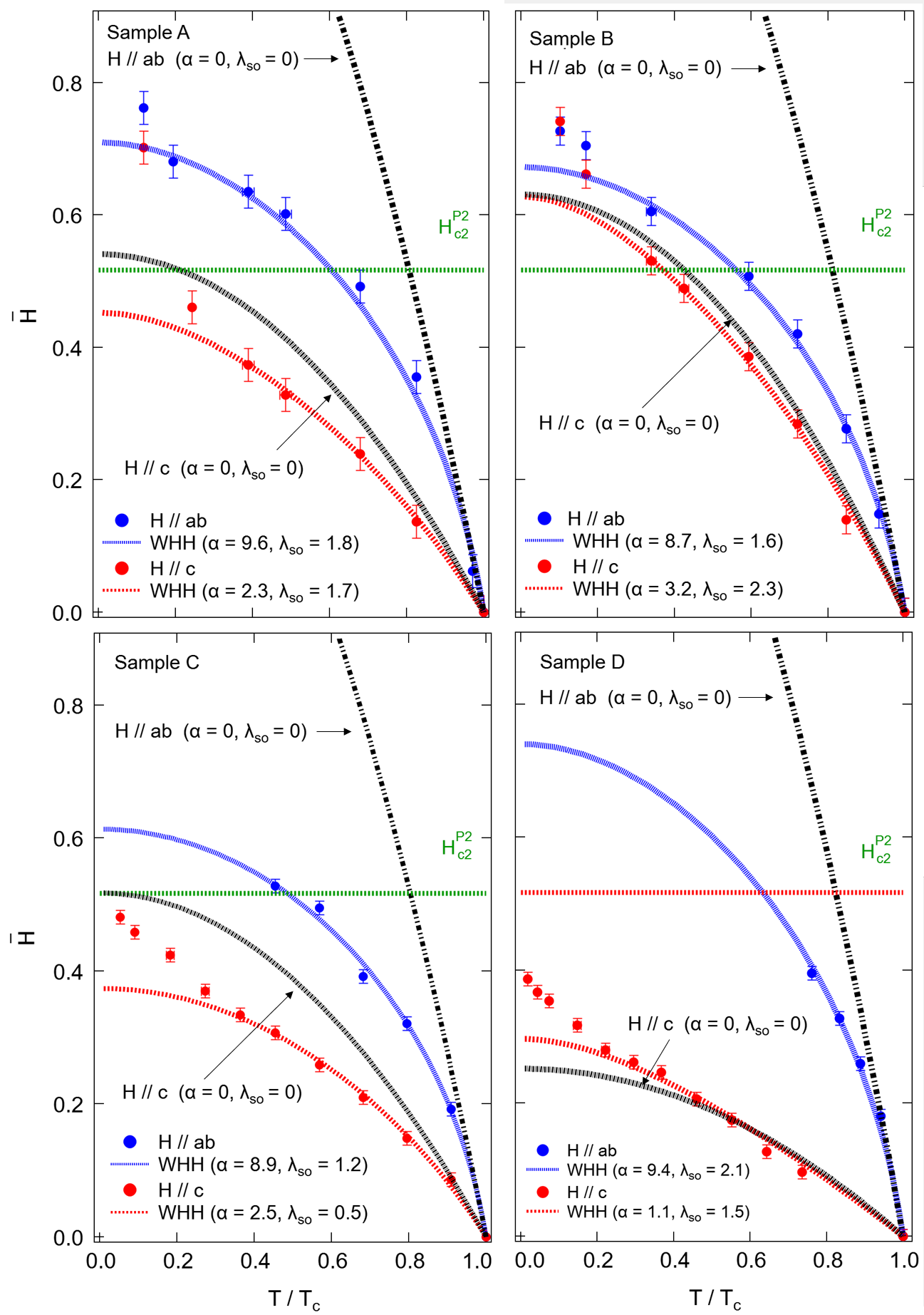}
      \caption{The dimensionless upper critical field $\bar{H} = \alpha h$ plotted as a function of the normalized temperature $t = T/T_{c}$ for samples (a) A, (b) B, (c) C, and (d) D, along with their corresponding WHH fits for $H \parallel ab$ and $H \parallel c$. The horizontal dashed line in each plot corresponds to the strongly coupled paramagnetic limit $\mu_{0}H_{c2}^{P2} = 1.86(1+\lambda_{el-ph})\: T_{c}$, while the ($\alpha = 0, \lambda_{so}=0$) curves demonstrate the case of strictly orbital-limitation and yield the results of Eq. 3. 
       }
      
      \label{fig:WHH}
  \end{figure*}

\newpage

\begin{figure}
    \centering
    \includegraphics[scale=0.25]{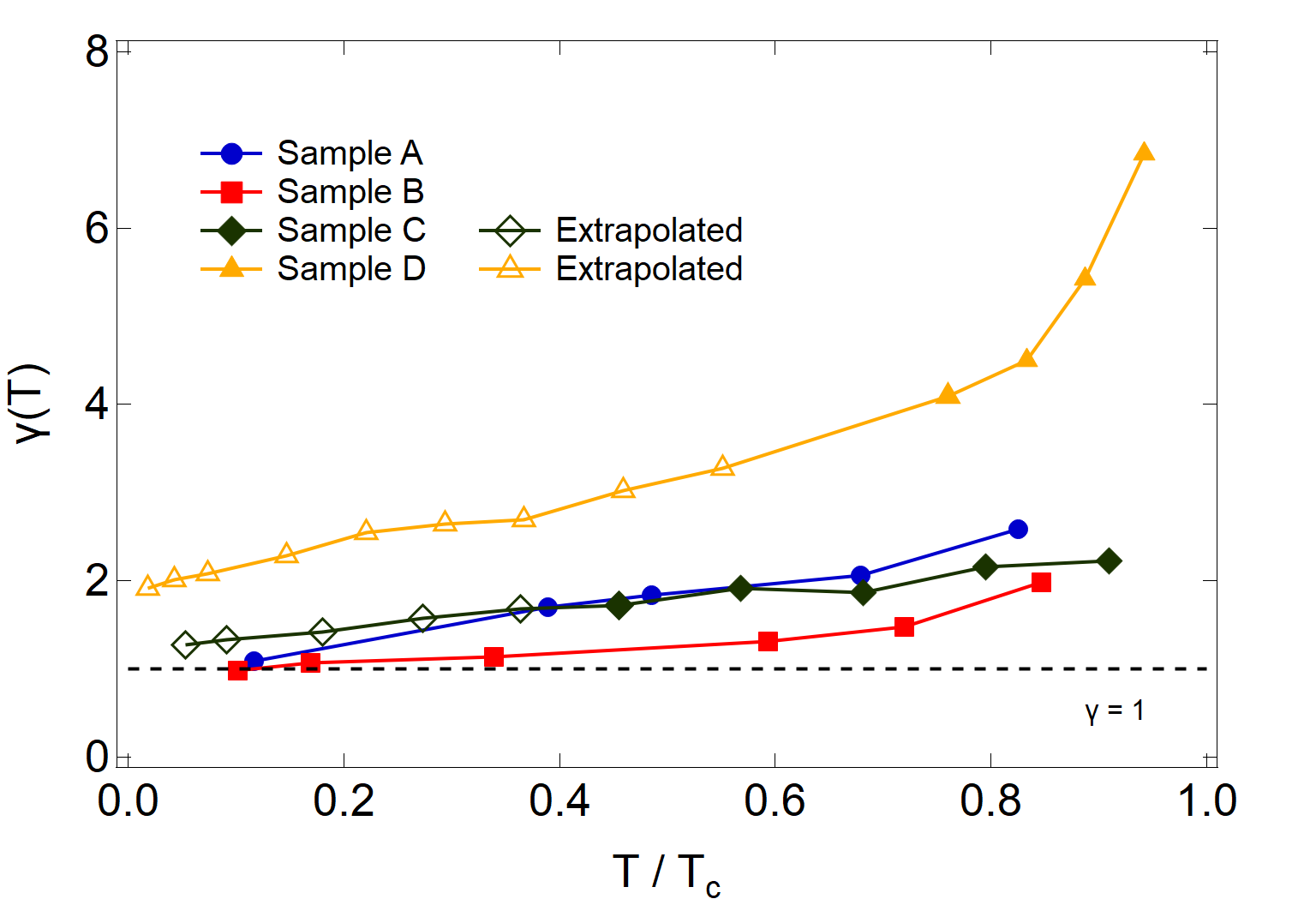}
    \caption{Temperature dependence of the upper critical field anisotropy parameter $\gamma = H_{c2}^{ab}(T)/H_{c2}^{c}(T)$. Filled data points are obtained from actual experimental results where $H_{c2}^{ab}$ and $H_{c2}^{c}$ were obtained at the same temperature. Low temperature, open (unfilled) points for sample C were calculated using WHH-predicted values of $H_{c2}^{ab}(T)$ and experimentally measured values of $H_{c2}^{c} (T)$. Open points for sample D were calculated using experimental (predicted) values of $H_{c2}^{ab}(T)$ ($H_{c2}^{c}(T)$)  above 20 K and predicted (experimental) values at and below 20 K. Lines are a guide to the eye.}
    \label{fig:Anisotropy}
\end{figure}

\newpage

\end{document}